\documentstyle[12pt]{article}
\newcommand{\be}{\begin{equation}}
\newcommand{\ee}{\end{equation}}

\begin{document}
\bibliographystyle{unsrt}

\begin{center}
{\Large {\bf Factoring and Fourier transformation with a Mach-Zehnder
interferometer}}

\vspace{3mm}

Johann Summhammer \footnote{e-mail: summhammer@ati.ac.at}

\vspace{3mm}

Atominstitut

Schuettelstr. 115, A-1020 Vienna, Austria

\end{center}

\begin{abstract}
The scheme of Clauser and Dowling (Phys. Rev. {\bf A 53}, 4587 (1996)) for
factoring $N$ by means of an $N$-slit interference experiment is translated into
an experiment with a single Mach-Zehnder interferometer. With dispersive phase
shifters the ratio of coherence length to wavelength limits the numbers that can
be factored. A conservative estimate permits $N \approx 10^7$. It is furthermore
shown, that sine and cosine Fourier coefficients of a real periodic function can
be obtained with such an interferometer.
\end{abstract}

\noindent
PACS: 03.65.Bz, 06.50.Mk, 07.60.Ly

\vspace{5mm}

Recently Clauser and Dowling (CD) have shown that factors of an integer $N$ can
be determined by simply measuring the peaks of the intensity distribution on the
screen behind a Young's $N$-slit arrangement \cite{CD}. This device
is distinct from the currently much debated proposals for quantum computation,
because it does not need the entanglement of several quantal degrees of freedom.
Therefore it is very immune against decoherence and could readily be implemented
with present technology. The drawback is that it will not exhibit the potential
increase in computational power characteristic of entanglement. Nevertheless,
the work of CD indicates that single particle interference arrangements have
useful applications beyond physical measurements.

The purpose of this note is to point out that the CD-proposal can be translated
into an experiment with only a single Mach-Zehnder interferometer. This will
enhance the flexibility of this calculational device. A further point is that
a Mach-Zehnder interferomter can also perform other computations, in particular
cosine and sine Fourier transformations.

Let us first focus on the work of CD. This proposal shows, that in a suitably
chosen central region on the detection screen behind an $N$-slit arrangement,
all intensity peaks are equal if, and only if, the quantity $n \equiv \lambda R
/ a^2$ is a factor of $N$, where $\lambda$ is the wavelength of the incident
radiation, $a$ is the center-to-center distance of the slits, and $R$ is the
distance between the slits and the screen. Furthermore, the Fraunhofer limit is
assumed ($R \gg a \gg \lambda$). Different values of $n$ can be tested by
adjusting any of the parameters, for instance $\lambda$. The
probability amplitude caused by slit $i$ at a point ${\bf r}$ on the screen is
given by $\psi_i({\bf r})$. The probability that the particle will hit this
point is therefore
\be
I({\bf r}) = \left| \sum_{i=1}^N \psi_i({\bf r}) \right|^2 = \sum_{i=1}^N
\left| \psi_i ({\bf r}) \right|^2 + 2 \sum_{i=1}^{N-1} \sum_{j=i+1}^N
Re \left[ \psi_i({\bf r}) \psi_j^* ({\bf r}) \right] .
\ee
For the idea to be presented here, it is important to notice that the essential
properties of $I({\bf r})$ can also be obtained in measurements with a series of
different {\it two-slit} experiments. This is so, because
quantum mechanical probabilities are the second power of probability amplitudes.
Therefore, the interference term of $I({\bf r})$, which is the double summation
on the right hand side, only contains products made up of just two factors.
And this feature is already present in the two-slit experiment.
(This would not be the case, if, say, probabilities were the third power of
probability amplitudes. Then the paradigm of quantum physics would only be fully
contained in a three slit experiment.) The analytic behavior
of the interference term of the $N$-slit experiment can therefore be recaptured,
when adding the {\it intensities} of all permutations of two-slit experiments
obtainable by covering $N-2$ of the $N$ slits. This procedure will not
reproduce the behavior of the mean intensity. But we can neglect the
mean intensity here, because it is only of statistical relevance.

Instead of doing several two-slit experiments with different distances of the
slits, it is of course more convenient to use a Mach-Zehnder interferometer and
vary the phase shift. A further advantage of a Mach-Zehnder interferometer is
that it has only two outputs, instead of the continuum of a screen. Then, with
an arrangement as shown in Fig.1, factoring can be achieved in the following
way: The phase shift $\chi$ is increased in discrete steps $2 \pi / n$, such
that $\chi_j = 2 \pi j / n$. For simplicity we assume that at each step one
particle is sent into the interferometer. We look at the outputs only when
$j=kN$, $k=1,2,3...$ . So at the $k^{th}$ observation the probability to
register the particle at detector $A$ is
\be
p_A(k) = \frac{1}{2} \left[ 1 + \cos \left( \frac{2\pi k N}{n} \right) \right] .
\ee
We have $p_A(k) = 1$ only if $n$ is a factor of N. Suppose we perform $n$
observations, starting with $k=1$ until $k=n$. Then the sum of the particles
registered in detector $A$ will be
\be
I_n = \sum_{k=1}^n p_A (k) = \frac{n}{2} + \frac{1}{2} \sum_{k=1}^n \cos \left(
\frac{2 \pi k N}{n} \right) .
\ee
For $n$ a factor of $N$ we have $I_n = n$, because then the phases in the cosine
terms will all be multiples of $2 \pi$. However, if $n$ is not a factor, the
cosine-terms of the summation will roughly cancel each other, and we will only
have $I_n \approx n/2$. To see this more clearly, we write $N/n = M +r$, where
$M$ is the largest integer for which $N/n > M$, and $r$ is the rational
remainder, $r= L/n$, with $L \in [1,2,....,n-1]$. Then the summation over
the cosine terms can be written as
\be
\sum_{k=1}^n \cos \left( \frac{2 \pi L}{n} k \right).
\ee
The phases of the cosines are now equidistantly spread in $n$ steps from the
value $2 \pi L / n$, which is less than $2\pi$, to $2 \pi L$, which is at most
$2 \pi (n-1)$, so that when reducing the phases to the interval $[0,2 \pi]$,
this interval will be used quite evenly.

For practical applications it is important to know what calculations are
possible. Since the best interferometers today are operated with light, it is
sufficient to look at the characteristics of such interferometers.
If we implement the phase shift in the usual manner by increasing the path
length difference of the two arms of the interferometer, the coherence
length of the incident radiation sets the limit to the largest number
$N$ that can be factored. According to eq.(3) the maximum phase shift to
be set in this scheme is $2 \pi N$. Assuming a wavelength of $\lambda = 500 nm$
and conservatively limiting the coherence length, given by $C = \lambda^2 /
\Delta \lambda$, where $\Delta \lambda$ is the standard deviation of the
wavelength distribution, to $C=5m$, numbers up to $N \approx 10^7$ can be
factored.

One also wants to know how long a calculation will take. When testing
whether $n$ is a factor of $N$,
one must set $nN$ different phase shifts. Since the largest factor to be checked
is of the order of $\sqrt{N}$, the longest check will take a time proportional
to $N^{3/2}$.

A further consideration is the maximum deviation permissible in the
phase increments $2 \pi/n$,
in order to ensure correct identification of the factors of $N$. Suppose
the actual increment is $2 \pi /(n+d)$, where $d$ is the deviation.
In the ideal case we have $d=0$, and when $n$ is a factor of $N$ the sum (3)
yields an intensity of $n$. Now this sum will still be substantially larger than
$n/2$ (which would indicate a non-factor), if the deviation $d$ is such, that
for the largest phase shift to be set (at $k=n$) we have $2\pi N -\frac{\pi}{2}$
rather than the ideal value $2\pi N$. Then the contribution of the last term of
the sum is zero, whereas all the others are still positive, so that the total
intensity will be roughly $n(\frac{1}{2} + \frac{1}{\pi})$. This limits the
permissible relative
deviation of the phase shift increments from their ideal value to $\left|
\frac{d}{n} \right| \le (4N)^{-1}$. For our example of above this would mean
$\left| \frac{d}{n} \right| \le 2.5 \times 10^{-8}$. Such accuracy can be
achieved in optical interferometers, when the lengthening of one arm relative to
the other is itself controlled interferometrically.

It is also possible to implement a kind of parallel computation with a setup as
illustrated in Fig.2. Here, the numbers $n_1$, $n_2$, ..., $n_7$ can almost
simultaneously be checked for being factors of $N$, thereby utilizing an
incident particle for more than just one computation. The phase shifters in the
various loops are simultaneously stepped up, but at the different increments $2
\pi /n_1$, $2 \pi /n_2$, ...,  $2 \pi /n_7$, respectively. Let us assume $n_i <
n_j$ for $i<j$. And for the sake of convenience we sum the intensity in a given
detector as needed for testing for the largest of the possible factors of the
loops involved. For instance, at detector $A$ (and also at $B$) we would sum
until $k=n_4$, thus
\be
I_A = \sum_{k=1}^{n_4} \frac{1}{8}
\left[ 1 + \cos \left( \frac{2 \pi k N}{n_1} \right) \right]
\left[ 1 + \cos \left( \frac{2 \pi k N}{n_2} \right) \right]
\left[ 1 + \cos \left( \frac{2 \pi k N}{n_4} \right) \right].
\ee
The expectation values for $I_A$, $I_B$ and $I_C + I_D$ (the latter also just
summed up to $n_4$) for the various possibilities are shown in
Table I. Clearly, the results permit unique identification of the
eight possible answers of interest.

\vspace{5mm}

{\bf Table I.} Dependance of intensities on property of "factor" or
"non-factor".

Factors are indicated by "F", non-factors by "$-$". Intensities in units of $n_4
/8$.

\vspace{5mm}

\begin{tabular}{|c|c|c||c|c|c|} \hline
$n_1$ & $n_2$ & $n_4$ & $I_A$ & $I_B$ & $I_C + I_D$ \\ \hline \hline
F & F & F & 8 & 0 & 0 \\       \hline
F & F & $-$ & 4 & 4 & 0 \\     \hline
F & $-$ & F & 4 & 0 & 4 \\     \hline
F & $-$ & $-$ & 2 & 2 & 4 \\   \hline
$-$ & F & F & 4 & 0 & 0 \\     \hline
$-$ & F & $-$ & 2 & 2 & 0 \\   \hline
$-$ & $-$ & F & 2 & 0 & 2 \\   \hline
$-$ & $-$ & $-$ & 1 & 1 & 2 \\ \hline
\end{tabular}

\vspace{5mm}

The situation is a little
bit more complicated for the interferometer loops following the lower arm of the
$n_1$-interferometer. We must account for the possibility that $n_1$ is a factor
of $N$. If we then measured the intensity at any of the detectors $E$ to $H$
after every $N$ phase increments, we would never detect a particle. Therefore it
is necessary to start the loops on the lower arm of $n_1$ with a delay of around
$n_1 / 2$ increments. Similar arguments apply to any of the interferometers
connected to an output which is dark when the phase shift is a multiple of $2
\pi$. For instance, testing whether $n_3$ and $n_6$ are factors of $N$ requires
that one waits for about $(n_1 + n_3)/2$ increments, and from then on sums the
intensities at $E$ (or at $F$) after every $N$ further increments. Then the
expectation values for $I_E$ for the eight possibilities (i.e., which of $n_1$,
$n_3$, $n_6$ is a factor of $N$) are analogous to those shown for $I_A$ in Table
I. However, the offset relative to the total number of phase increments to be
gone through is small, such that one can still speak of essentially parallel
computation. But, of course, as this is a device not exploiting the entanglement
of quantum systems, the number of loops, or the total amount of time, needed for
factoring $N$ rises polynomially with $N$, rather than just logarithmically.

Let us now turn to how Fourier transformations can be performed by means of an
interferometer. Specifically, it is possible to obtain
the cosine and sine Fourier coefficients of a real valued positive
function $f(t)$, which is periodic with the period $\tau$, and where $t$ denotes
time. For
this it is necessary that the intensity incident on the
interferometer varies in time proportional to $f(t)$, and that the phase
shift $\chi$  applied between the two paths of the interferometer increases
linearly with time, e.g.
\be
\chi(t) = \frac{2\pi m t}{\tau},
\ee
where $m$ is an integer.
In order to avoid problems of the change of energy of the particles when
experiencing a time-dependent phase shift, we limit ourselves to slow rates,
$m/\tau \ll E/h$, where $E$ is the energy of the particles and $h$ is Planck's
constant.
Then the difference of the intensities at detectors $A$ and $B$ integrated
over one period $\tau$ is
\be
I_A - I_B = c \int_0^\tau f(t) \cos \left( \frac{2 \pi m t}{\tau} \right) dt ,
\ee
where $c$ is the proportionality constant between $f(t)$ and the incident
intensity. As can be seen, $I_A -I_B$ is proportional to the $m^{th}$ cosine
Fourier coefficient of $f(t)$. In order to obtain the sine Fourier coefficients,
one only needs to add a constant phase shift of $-\pi/2$ to $\chi(t)$.

Note that this calculational procedure is different from the experimental
method employed in Fourier spectroscopy, which has been used in astronomy
since the days of Michelson and which has recently also been used in electron
\cite{Hasselbach}
and neutron \cite{Rauch} interferometry. In Fourier spectroscopy one can measure
the spectral
distribution of the incident radiation, if this distribution is symmetric about
the mean, but one gets no phase information.

\vspace{20mm}

{\Large {\bf Figure Captions}}

\vspace{5mm}

\noindent
{\bf Fig.1:} Ideal Mach-Zehnder interferometer with semitransparent mirrors for
beam splitting at the entrance and beam superposition at the exit. A phase
shifter introduces the additional phase $\chi$ between the two paths. Particles
are registered in detectors $A$ and $B$.

\vspace{5mm}

\noindent
{\bf Fig.2:} Series connection of several Mach-Zehnder interferometers for
almost simultaneous determination whether the integers $n_1$, ... , $n_7$ are
factors of $N$. The particles are registered at detectors $A$ to $H$.

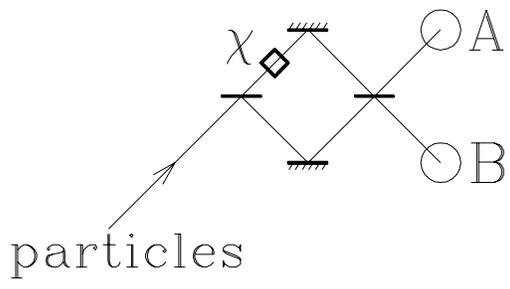
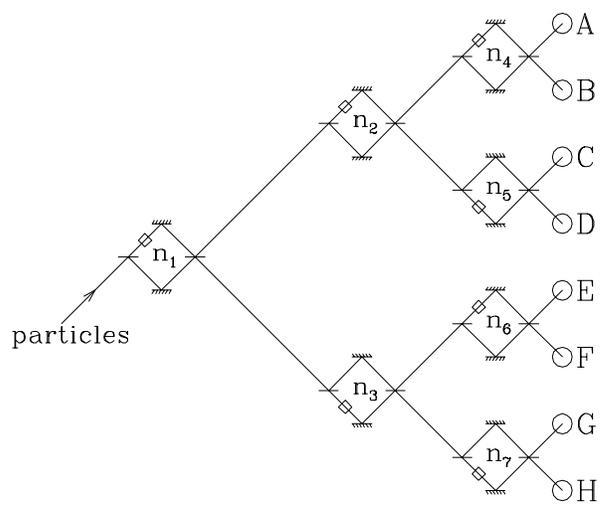
\end{document}